          [2001/04/25 1.1 (PWD)]
\documentclass[a4paper,twocolumn]{esapub} 
\usepackage{natbib,graphicx,amssymb}

\title{Multi-wavelength INTEGRAL NEtwork (MINE) observations of the
microquasar GRS\,1915+105}
\author{Ya\"el Fuchs}
\affil{Service d'Astrophysique, CEA/Saclay, Orme des Merisiers b\^at. 709, 91191 Gif sur Yvette cedex, France}
\author{J. Rodriguez$^{1,}$}
\affil{Integral Science Data Centre, Chemin d'Ecogia, 16, CH-1290 Versoix, Switzerland}
\author{S. E. Shaw$^{2,}$}
\affil{School of Physics and Astronomy, University of Southampton, Southampton, SO17 1BJ UK}
\author{P. Kretschmar$^{2,}$}
\affil{Max-Planck-Institut fuer Extraterrestrische Physik, Giessenbachstrasse, 85748 Garching, Germany}
\author{M. Rib\'o$^{1}$, S. Chaty$^{1,}$}
\affil{Universit\'e Paris 7, 2 place Jussieu, 75005 Paris, France}
\author{I.~F.~Mirabel$^{1,}$}
\affil{Instituto de Astronom\'\i a y F\'\i sica del Espacio / CONICET, 
cc67, suc 28. 1428 Buenos Aires, Argentina}
\author{V. Dhawan}
\affil{National Radio Astronomy Observatory, Socorro, NM 87801, USA}
\author{G. G. Pooley}
\affil{Mullard Radio Astronomy Observatory, Cavendish Laboratory, Madingley Road, Cambridge CB3 0HE, UK}
\author{I. Brown}
\author{R. Spencer}
\affil{University of Manchester, Jodrell Bank Observatory, Macclesfield, Cheshire SK11 9DL}
\author{D. C. Hannikainen}
\affil{Observatory, University of Helsinki, PO Box 14, FIN-00014 , Finland}

\begin{document}


\maketitle

\begin{abstract}
We present the international collaboration MINE (Multi-lambda Integral
NEtwork) aimed at conducting multi-wavelength observations of X-ray
binaries and microquasars simultaneously with the {\it INTEGRAL}
$\gamma$-ray satellite. We will focus on the 2003 March--April
campaign of observations of the
peculiar microquasar GRS\,1915+105 gathering radio, IR and X-ray
data. The source was observed 3 times in the plateau state, before and
after a major radio and X-ray flare. 
   It showed strong steady optically thick radio emission corresponding to 
   powerful compact jets resolved in the radio images, 
   bright near-infrared emission, a strong QPO at 2.5\,Hz in the X-rays
   and a power law dominated spectrum without cutoff in the 3--300\,keV range. 
We
compare the different observations, their multi-wavelength light
curves, including JEM-X, ISGRI and SPI, and the parameters deduced
from fitting the spectra obtained with these instruments on board
{\it INTEGRAL}.
\end{abstract}

\section{Introduction}
	Microquasars are X-ray binaries that produce relativistic jets
	and thus appear as miniature replicas of distant quasars
	\citep{mirabelrodriguez99}.
	Their emission
	spectra, variable with time, range from the radio to the
	$\gamma$-ray wavelengths.
	We present here the first multi-wavelength campaign on
	GRS\,1915+105 involving the {\it INTEGRAL} satellite (3\,keV--10\,MeV).
	This campaign was conducted by the MINE
	(\mbox{Multi-$\lambda$} {\it INTEGRAL} NEtwork, see
	{\sf http://elbereth.obspm.fr/$\sim$fuchs/mine.html}) international
	collaboration aimed at performing multi-wavelength
	observations of galactic X-ray binaries simultaneously with
	{\it INTEGRAL}.

\section{GRS\,1915+105}
	The microquasar GRS\,1915+105 is extremely variable at all
	wavelengths (see \citealt{fuchs03} for a review).  It hosts the
	most massive known stellar mass black hole of our Galaxy with
	M$=14.0\pm4.4$\,M$_{\odot}$ 
	\citep{greiner01Nat, harlaftis04}.  
	In the radio 
	it can show superluminal ejections at arcsec scales
	\citep{mirabelrodriguez94}
	leading to a maximum distance of 11.2$\pm$0.8\,kpc 
	\citep{fender99}, 
	and  
	compact jets at milli-arcsecond scales 
	(= a few tens of AU, \citealt{dhawan00}).
	 
\begin{figure}[!ht]
   \centering
   \includegraphics[angle=-90,width=\columnwidth]{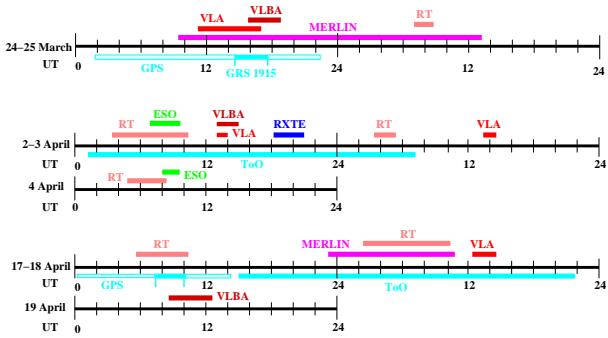}
      \caption{Viewgraph of the observing campaign in spring
      2003, indicating the dates, time and involved observatories (GPS
      = Galactic Plane Survey of {\it INTEGRAL}).}
       \label{figcamp}
\end{figure}

	We conducted a multi-wavelength observation campaign of
	GRS\,1915+105 on 
	March 24--25, April 2--3 and April 17--18 2003 (see
	Fig.~\ref{figcamp}) corresponding to {\it INTEGRAL}
	revolutions 54, 57 and 62, respectively.  Here we focus only on
	the April observations when ToO (Targets of Opportunity) were
	triggered by the MINE collaboration.  This (nearly)
	simultaneous campaign involved the VLA, the VLBA, MERLIN and
	the Ryle Telescope (RT) in radio, the ESO/NTT in IR, {\it
	RXTE} and {\it INTEGRAL} in X/$\gamma$-rays. More details and
	description of the April 2 observations can be found in 
	\citet{fuchsmine03}.

\begin{figure}[!ht]
   \centering
   \includegraphics[width=\columnwidth]{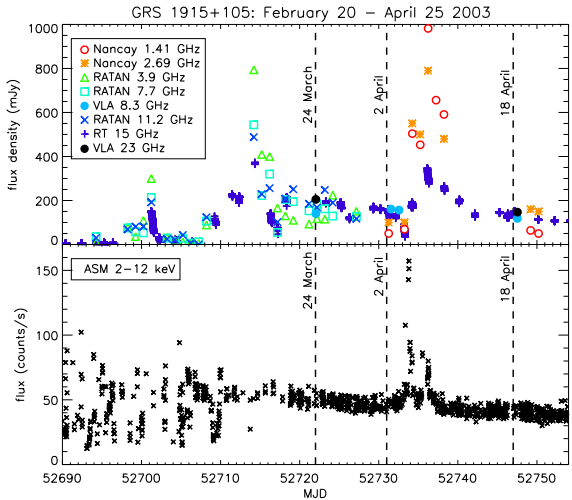}
   \vspace*{-0.3cm}
      \caption{Radio and X-ray flux monitoring of GRS\,1915+105 in
      March--April 2003.  Radio: Nan\c cay at 1.41 \& 2.69~GHz, RATAN at
      3.9, 7.7 \& 11.2~GHz, VLA at 8.4 \& 22~GHz, RT at
      15\,GHz. X-ray: quick-look results provided by the ASM/RXTE
      team. The dashed lines indicate the dates of our INTEGRAL and
      simultaneous multi-wavelength observations.  The RATAN flux
      densities were kindly provided by Sergei Trushkin.  The Nan\c cay
      flux densities were kindly provided by Harry Lehto and Emilios
      Harlaftis.
}
       \label{figclgen}
\end{figure}

\begin{figure}[!ht]
   \centering
   \includegraphics[width=\columnwidth]{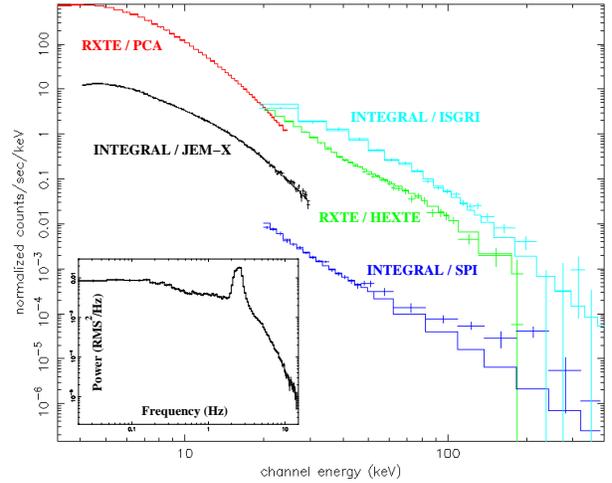}
   \vspace*{-0.3cm}
      \caption{X/$\gamma$-ray spectra and fit of
   GRS\,1915+105 measured with {\it RXTE} \& {\it INTEGRAL}
    on April 2, 2003 (rev.\,57). 
   The structures at $E$$>$50\,keV in
   the SPI spectrum are instrumental background lines not adequately corrected.
   The PCA power density spectrum (inset)
   shows a clear QPO at 2.5\,Hz.}
       \label{figspec}
\end{figure}

\begin{figure}[!ht]
   \centering
   \includegraphics[width=\columnwidth]{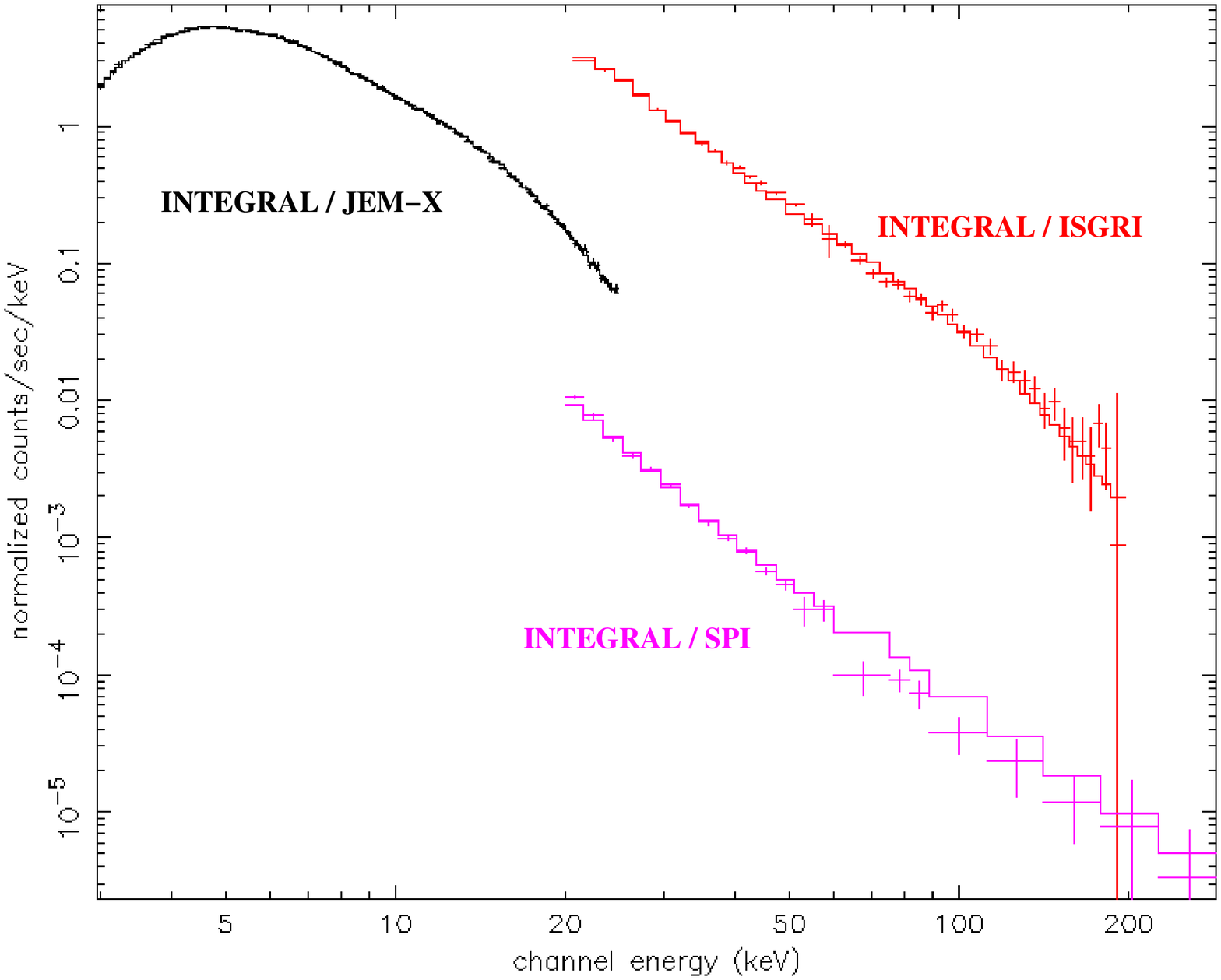}
   \vspace*{-0.3cm}
      \caption{X/$\gamma$-ray spectra and fit of
   GRS\,1915+105 measured with {\it INTEGRAL} on April 17--18, 2003 
   (rev.\,62).}
       \label{figspec18av}
\end{figure}

\begin{figure}[!ht]
   \centering
   \includegraphics[width=\columnwidth]{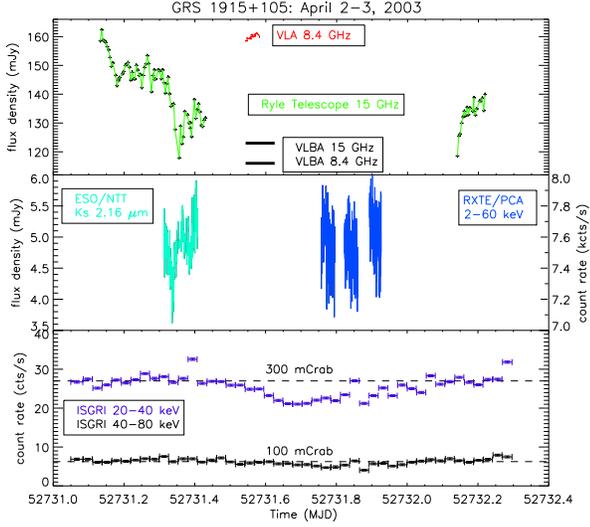}
   \vspace*{-0.3cm}
      \caption{Light curves of the multi-wavelength observations on
      April 2--3, 2003 (rev.\,57) involving INTEGRAL/ISGRI \& SPI (not
      plotted), RXTE/PCA \& HEXTE (not plotted), ESO/NTT, RT, VLA and
      VLBA (total flux densities spanning the observing time).
}
       \label{figcl2av}
\end{figure}

\begin{figure}[!ht]
   \centering
   \includegraphics[width=\columnwidth]{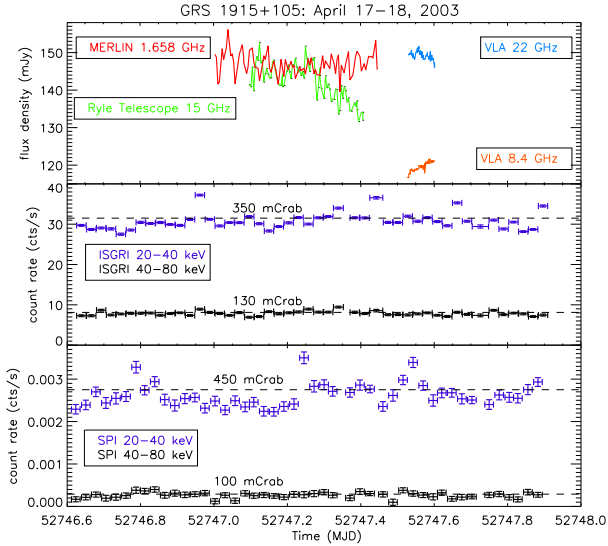}
   \vspace*{-0.3cm}
      \caption{Light curves of the multi-wavelength observations on 
	April 17--18, 2003 (rev.\,62) involving INTEGRAL/ISGRI \& SPI, 
	RT, MERLIN, VLA and VLBA 
	(total flux densities spanning the observing time).
	The difference between the flux levels of ISGRI and SPI is due 
	to calibration.}
       \label{figcl18av}
\end{figure}

\begin{figure}[!ht]
   \centering
   \includegraphics[width=\columnwidth]{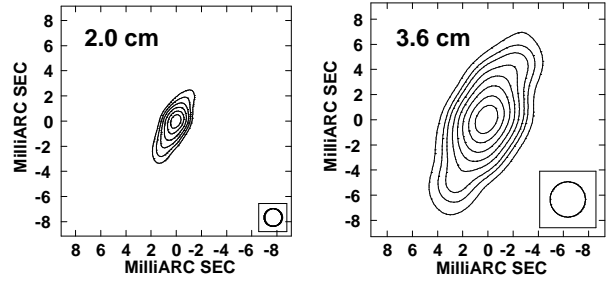}
   \vspace*{-0.3cm}
      \caption{VLBA images at 2.0 \& 3.6\,cm on April~2, 2003 showing
      the compact jet. 
	1\,mas = 12\,AU at 12\,kpc.} 
       \label{figjet}
\end{figure}

\begin{figure}[!ht]
   \centering
   \includegraphics[angle=-90,width=\columnwidth]{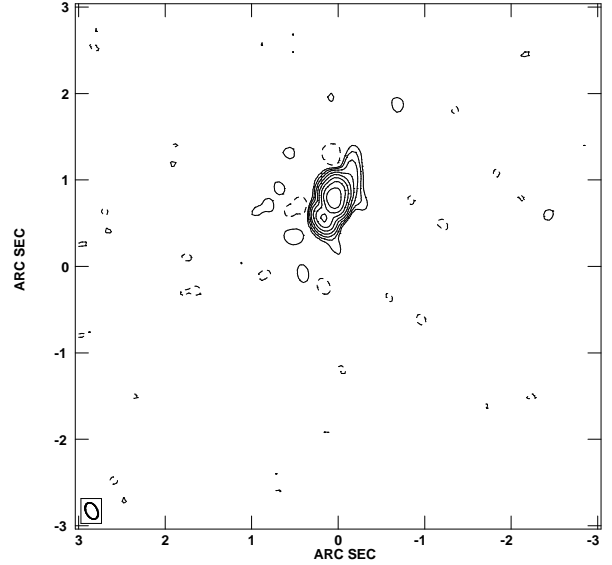}
   \vspace*{-0.3cm}
      \caption{MERLIN 18\,cm (1.658\,GHz) image of GRS\,1915+105 
	on April 17--18, 2003.}
       \label{figmerlin}
\end{figure}

	Fig.~\ref{figclgen} shows the general radio and X-ray
	behaviour of GRS\,1915+105 during our observing campaign.
	After a period of X-ray oscillations (end of February --
	beginning of March) giant radio flares occured around MJD
	52714 (March 15) and MJD 52734 (April 5). The latter may be a
	double flare since there are two X-ray flares in the {\it
	RXTE}/ASM light curve, and it probably corresponds to major
	superluminal ejections.
	Despite these flaring episodes, our
	observations took place during the \emph{plateau}
	state of GRS\,1915+105 \citep{fender99,kleinwolt02} 
	i.e.\@ quasi-steady
	 {\it RXTE}/ASM (2--12\,keV) flux
	$\sim$50\,cts/s and  high steady radio level 
	($>$100\,mJy).
	This \emph{plateau} state was observed on several past occasions in
	GRS\,1915+105 (see e.g. Fig.\,1 of \citealt{muno01}) and was
	also called the radio loud low/hard X-ray state \citep{muno01}
	and type II state \citep{trudolyubov01}. It also corresponds
	to the $\chi_1$ and $\chi_3$ X-ray classes of
	\citet{belloni00} who used PCA color-color diagram, and indeed
	our April 2 observation appears as the $\chi_1$ class when
	plotted in such a diagram.

	On April\,2 (Fig.~\ref{figspec}) \& 18 (Fig.~\ref{figspec18av})
	the high energy emission 
	shows a power law dominated spectrum ($>$60\% at
	3--20\,keV) with a photon index $\Gamma$=2.9 and 2.75, respectively. 
	This state is much softer 
	than the classical low/hard state of the other BH binaries 
	and is closer to the very high or intermediate states
	\citep{mcclintock04}. 
	The {\it INTEGRAL} observations
	show that this power law spectrum extends up to 300\,keV
	without any cutoff during this \emph{plateau} state, consistent with
	the observations with {\it CGRO}/OSSE 
	\citep{zdziarski01}.
%
	The estimated luminosity on April 2 is 
	$\sim$\,$7.5$\,$\times$\,$10^{38}$\,erg\,s$^{-1}$, 
	using the {\it RXTE}/PCA (2--30\,keV) which overestimates the flux 
	by about 16\%. 	
	On April 18 the luminosity is 
	$\sim$\,$2.8$\,$\times$\,$10^{38}$\,erg\,s$^{-1}$ using {\it
	INTEGRAL}/JEM-X (3--30\,keV).
	These values correspond to 40\% and 16\%, 
	respectively,
	of the Eddington luminosity for a $14M_{\odot}$ black hole.
	As shown in Fig.~\ref{figspec}, a very
	clear Quasi-Periodic Oscillation (QPO) at 2.5\,Hz with a 14\%
	rms level was observed in the {\it RXTE}/PCA signal, 
	which is
	consistent with the previous observations of the \emph{plateau} state
	(see also \citealt{rodriguez04}).

	The detailed light curves of our observations on April~2--3
	(Fig.~\ref{figcl2av}) and April~17--18 (Fig.~\ref{figcl18av}) 
	show nearly quiet flux densities
	when compared to flares or oscillations, with variations
	$\lesssim$20\% at all wavelengths
	(the peaks in the ISGRI and SPI light curves are not real).
	The most interesting phenomenon is on April~2, a moderate $\sim$25\%
	decrease (from 4.9 to 3.6\,mJy) in the $K_{\mathrm s}$ flux density 
	lasting 20\,min which precedes by 31\,min a $\sim$20\% decrease in
	the RT signal (from 145 to 118\,mJy) lasting 48\,min.  This 
	may be due to instabilities in the jet
	inducing an
	immediate synchrotron response in the IR and the delay being
	due to the time for the material along the jet to become
	optically thin to the radio emission \citep{mirabel98}. 
	The radio and near-IR flux densities are high in both observations.

	The VLBA high resolution images on April 2
	(Fig.~\ref{figjet}) show the presence of a compact
	radio jet with a $\sim$7--14\,mas length (85--170\,AU at
	12\,kpc).
	These jets are also observed in the radio images on March~24 
	and April~19.
	The optically thick synchrotron emission from this jet is
	responsible for the high radio levels observed during the
	\emph{plateau} state of GRS\,1915+105 (see Fig.~\ref{figclgen},
	Fig.~\ref{figcl2av} and Fig.~\ref{figcl18av}).  On April 18
	the MERLIN image (Fig.~\ref{figmerlin}) shows a radio extension
	of $\sim$0.3$''$. This extension is likely
	the trace of a superluminal ejection occurred around April~4
	(corresponding to the giant radio flare of April~5 on
	Fig.~\ref{figclgen}) since it would correspond to a mean
	velocity of $\sim$\,21\,mas/d compatible with the previous
	observations (\citealt{mirabelrodriguez99}).

\begin{figure}[!tbp]
\includegraphics[width=\columnwidth]{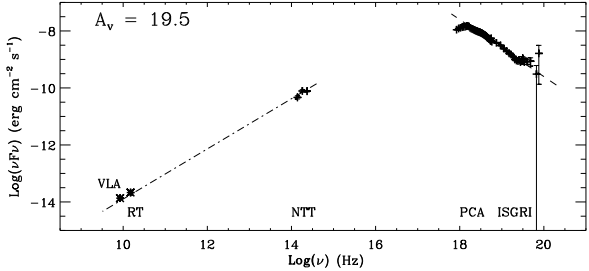}
\vspace*{-0.7cm}
\caption{
   Spectral energy distribution of GRS\,1915+105 on April~2. The
   near-IR flux densities were dereddened with $A_V$\,=\,19.5 and
   using eq.\,1 of \citet{cardelli89}. The dot-dashed line illustrates
   the optically thick synchrotron emission from the jet as a power
   law with $F_\nu\propto\nu^{-0.12}$. The dashed line illustrates a
   power law with a photon index $\Gamma$=3. Error bars not shown are
   smaller than the symbol size.  }
\label{figsed}
\end{figure}

	On April\,2 the source was fairly bright in near-IR 
	with an excess of 75--85\%
	in the $K_{\mathrm s}$-band compared to the $K$=14.5--15\,mag.\@
	of the K-M giant donor star of the binary.
	According to the spectral energy distribution (Fig.~\ref{figsed}), 
	this IR excess is 
	compatible with a strong contribution from
	the synchrotron emission of the jet extending from the radio
	up to the near-IR. 
	Different components, however, contribute to the IR
	in addition to the jet,
	such as the donor-star, the external part of the
	accretion disc or free-free emission.

\section{Conclusions and Prospects}
	Here for the first time, we observed simultaneously all the
	properties of the \emph{plateau} state of GRS\,1915+105\,: 
	a powerful compact
	radio jet, responsible for the strong steady radio emission and
	probably for a significant part of the bright near-IR
	emission, as well as a QPO (2.5\,Hz) in the X-rays 
	and a power law dominated X-ray spectrum with
	a $\Gamma$$\sim$3 photon index up to at least
	300\,keV. Forthcoming works will study
	detailed fits of the X-ray spectra, 
	to determine for example
	whether this power law is due to an inverse Compton
	scattering of soft disc photons 
	on the base of the compact jet 
	or not.
	In order to better
	understand the unusual behaviour of GRS\,1915+105, we need to
	carry out similar simultaneous broad-band campaigns during the
	other states, in particular during the sudden changes 
	 that correspond to powerful relativistic
	ejection events.

\section*{Acknowledgments}
  Y.F. and J.R. acknowledge financial support from the CNES.
  M.R. acknowledges support from a Marie Curie individual fellowship under
contract No.\,HPMF-CT-2002-02053.
  D.C.H. acknowledges the Academy of Finland for financial support.
  The National Radio Astronomy Observatory is a facility of
the National Science Foundation operated under cooperative agreement by
Associated Universities, Inc.
  Based on observations collected at the European Southern Observatory,
Chile (ESO N$^\circ$ 071.D-0073).
  We thank the ASM/RXTE team for providing the quick-look results.
  We thank Sergei Trushkin for the RATAN data, 
and Harry Lehto and Emilios Harlaftis for the Nan\c cay data.
    \bibliographystyle{aa}
\bibliography{ref1915-munich}

 \end{document}